# Extension and exact realization of the Heller's derivative rule

Hashim A. Yamani[a] and Abdulaziz D. Alhaidari[b]

[a] *King Abdullah City for Atomic and Renewable Energy, P. O. Box 22022, Riyadh 11451, Saudi Arabia*
[b] *Saudi Center for Theoretical Physics, Jeddah, Saudi Arabia*

**Abstract**. The accuracy of the Heller's derivative rule to calculate the numerical weights associated with discretized energy spectrum is enhanced by Broad's extension which adds ($N-1$) more interpolating points to the original $N$ points. The extension scheme is then used to show how to realize the rule without any approximation.

The idea of using a finite square-integrable ($L^2$) basis functions to perform scattering calculations was bold on two counts [1-6]. First, the $L^2$ basis is usually a bound-state-like tool usually reserved for structure calculation. It is thought to be counter-intuitive for this basis to represent non-square-integrable wavefunctions. Second, the finite nature of the basis destroys the needed analytic properties of important scattering quantities such as the S-matrix or the Green's function. This results from the fact that the continuous scattering spectrum is rendered discrete by the use of finite basis (See for example [6] for discussion of this point]).

It is now an established fact that the use of finite as well as complete $L^2$-basis functions proved to be very successful and efficient calculational tools with rich mathematical underpinnings. One such tool is the Heller's ansatz [7,8], which gives a clear and straightforward rule of how to approximate an integral over the continuous energy spectrum by a sum over discrete energies. More explicitly, if $H$ is the scattering Hamiltonian and $\bar{H}$ is the $N \times N$ matrix representation of $H$ in the finite basis $\{|\phi_n\rangle\}_{n=0}^{N-1}$ having discrete energy eigenvalues $\{\varepsilon_\mu\}_{\mu=0}^{N-1}$, then we have the following integral approximation

$$\int_0^\infty F(\varepsilon)\, d\varepsilon \cong \sum_{\mu=0}^{N-1} \omega_\mu^d F(\varepsilon_\mu). \qquad (1)$$

The procedure to specify the set of $N$ numbers $\{\omega_\mu^d\}_{\mu=0}^{N-1}$ is as follows:

(a) Find a function $\zeta(x)$ that interpolate the sorted eigenvalues $\{\varepsilon_\mu\}_{\mu=0}^{N-1}$ in the index $\mu$ such that
$$\zeta(\mu) = \varepsilon_\mu. \qquad (2)$$

(b) Calculate the derivative weights $\{\omega_\mu^d\}_{\mu=0}^{N-1}$ simply as
$$\omega_\mu^d = \left(\frac{d\zeta(x)}{dx}\right)_{x=\mu}. \qquad (3)$$

From the outset, it is useful to note that the $\{\omega_\mu^d\}_{\mu=0}^{N-1}$ are different from the usual quadrature approximation scheme for energy integrals where the abscissas $\{\varepsilon_\mu\}_{\mu=0}^{N-1}$ and the numerical weights $\{\omega_\mu\}_{\mu=0}^{N-1}$ are derivable from a positive density function $\rho(\varepsilon)$ so that [9]

$$\int_0^\infty \rho(\varepsilon) F(\varepsilon)\, d\varepsilon \cong \sum_{\mu=0}^{N-1} \omega_\mu F(\varepsilon_\mu). \tag{4}$$

We see that actually

$$\omega_\mu^d = \omega_\mu / \rho(\varepsilon_\mu). \tag{5}$$

The one often cited example where the Heller rule yields an exact result is the case of the Chebychev polynomials [8,9]. In this case

$$\varepsilon_\mu = -\cos\left(\frac{\mu+1}{N+1}\right)\pi, \quad ;\mu = 0,1,\ldots,N-1.$$

Naturally, then

$$\zeta(x) = -\cos\left(\frac{x+1}{N+1}\right)\pi.$$

Thus the rule gives

$$\omega_\mu^d = \frac{\pi}{N+1}\sin\left(\frac{\mu+1}{N+1}\right)\pi \quad ;\mu = 0,1,\ldots,N-1.$$

This result is in exact correspondence with Chebychev system where

$$\rho(\varepsilon) = \sqrt{1-\varepsilon^2}, \quad \text{and}$$

$$\omega_\mu = \frac{\pi}{N+1}\sin^2\left(\frac{\mu+1}{N+1}\right)\pi \quad ;\mu = 0,1,\ldots,N-1.$$

For other cases where no clear interpolation scheme is known, the rule is introduced by Heller as an approximation. In this letter, we show that Heller's ansatz has strong theoretical support. The first step in that direction was taken by Broad who carried out an analysis to mathematically justify and subsequently improve the accuracy of the Hellr's rule by essentially adding more interpolation points to the original $N$ points $\{\varepsilon_\mu\}_{\mu=0}^{N-1}$ [10]. Broad considered the finite Green's function matrix elements $g_{00}(z)$ which he wrote explicitly as

$$g_{00}(z) = \frac{\prod_{\nu=0}^{N-2}(z-\hat{\varepsilon}_\nu)}{\prod_{\mu=0}^{N-1}(z-\varepsilon_\mu)}, \tag{6}$$

where $\{\hat{\varepsilon}_\nu\}_{\nu=0}^{N-2}$ is the set of $(N-1)$ eigenvalues of the matrix abbreviated from $\bar{H}$ by deleting the first row and first column [11]. In the limit as $N \to \infty$ when the basis set becomes complete, it approaches the exact Green's function matrix element $G_{00}^{(+)}(z)$ associated with the scattering Hamiltonian $H$. Broad's significant contribution is in showing that the interpolating function $\zeta(x)$, or rather its inverse $\zeta^{-1}(\varepsilon)$, has the form

$$\tan\left[\pi\zeta^{-1}(\varepsilon)\right] = \frac{\mathrm{Im}\,G_{00}^{(+)}(\varepsilon)}{\mathrm{Re}\,G_{00}^{(+)}(\varepsilon) - g_{00}(\varepsilon)}. \tag{7}$$

We note immediately that this form already contains the Heller's interpolation condition of Eq. (2) since at the poles $\{\varepsilon_\mu\}_{\mu=0}^{N-1}$ of $g_{00}(z)$, $\tan\left[\pi\zeta^{-1}(\varepsilon)\right]$ vanishes, giving

$$\zeta^{-1}(\varepsilon_\mu) = \mu, \text{ an integer}.$$

Broad further argued that at the extra $(N-1)$ points $\{\hat{\varepsilon}_\nu\}_{\nu=0}^{N-2}$ which are the zeros of $g_{00}(z)$, we have

$$\tan\left[\pi\zeta^{-1}(\hat{\varepsilon}_\nu)\right] = \frac{\mathrm{Im}\,G_{00}^{(+)}(\hat{\varepsilon}_\nu)}{\mathrm{Re}\,G_{00}^{(+)}(\hat{\varepsilon}_\nu)} = \tan\left[\arg G_{00}^{(+)}(\hat{\varepsilon}_\nu)\right].$$

With $(2N-1)$ points the ansatz now reads as follows:
  (a) Choose the arg of $G_{00}^{(+)}(\varepsilon)$ to lie in the interval $[0,\pi]$ and, utilizing the fact that that two sets of energies $\{\varepsilon_\mu\}_{\mu=0}^{N-1}$ and $\{\hat{\varepsilon}_\nu\}_{\nu=0}^{N-2}$ interweave, sort them according to the scheme

$$\varepsilon_0 < \hat{\varepsilon}_0 < \varepsilon_1 < \hat{\varepsilon}_1 < \cdots < \hat{\varepsilon}_{N-3} < \varepsilon_{N-2} < \hat{\varepsilon}_{N-2} < \varepsilon_{N-1}$$

  (b) Augment the Heller's interpolation condition of equation (2) by the conditions

$$\zeta^{-1}(\hat{\varepsilon}_\nu) = \nu + \frac{1}{\pi}\arg\left[G_{00}^{(+)}(\hat{\varepsilon}_\nu)\right]; \quad \nu = 0,1,\ldots,N-2. \tag{8}$$

for a total of $(2N-1)$ interpolation points.

We naturally expect that as a result of the Broad's proposal, these many points pin down the function $\zeta(x)$ more accurately and consequently lead to a more accurate estimate of derivative weights $\{\omega_\mu^d\}_{\mu=0}^{N-1}$. In fact for cases where $G_{00}^{(+)}(\varepsilon)$ is exactly known, we have found [6] that Broad's extension indeed produce more accurate values of $\{\omega_\mu^d\}_{\mu=0}^{N-1}$. However, an accurate knowledge of $G_{00}^{(+)}(\varepsilon)$ is usually not available since it assumes knowledge of the system beyond knowledge of the properties of the finite Hamiltonian matrix $\bar{H}$.

We propose to show that the J-matrix approach [12-15] provides a way around this challenge while at the same time builds on the significant step taken by Broad. In this approach, we work in a complete basis set $\{|\phi_n\rangle\}_{n=0}^{\infty}$ in which the representation of reference Hamiltonian $H^0$ is tridiagonal and the coefficients of the associated asymptotically sine-like $|S(\varepsilon)\rangle$ and cosine-like $|C(\varepsilon)\rangle$ solutions of the reference problem are known,

$$|S(\varepsilon)\rangle = \sum_{n=0}^{\infty} s_n(\varepsilon)|\phi_n\rangle, \tag{9a}$$

$$|C(\varepsilon)\rangle = \sum_{n=0}^{\infty} c_n(\varepsilon)|\phi_n\rangle. \tag{9b}$$

Furthermore, the scattering potential $V$ is restricted to the subspace spanned by the finite basis $\{|\phi_n\rangle\}_{n=0}^{N-1}$, so that

$$H_{nm} = \begin{cases} H^0_{nm} + V_{nm} &, 0 \leq n, m \leq N-1 \\ H^0_{nm} &, \text{otherwise} \end{cases} \qquad (10)$$

In this setup, it has been previously worked out that [6]

$$G^{(\pm)}_{00}(\varepsilon) = g_{00}(\varepsilon) - g_{0,N-1}(\varepsilon) J_{N-1,N}(\varepsilon) \left( R^\pm_N(\varepsilon) / \Delta^\pm_N(\varepsilon) \right) g_{N-1,0}(\varepsilon), \qquad (11)$$

where

$$J_{nm}(\varepsilon) = \langle \phi_n | (H^0 - \varepsilon) | \phi_m \rangle,$$

$$g_{nm}(\varepsilon) = \langle \phi_n | (\bar{H} - \varepsilon)^{-1} | \phi_m \rangle = \sum_{\mu=0}^{N-1} \frac{\Gamma_{n\mu} \Gamma_{m\mu}}{\varepsilon_\mu - \varepsilon},$$

$$R^\pm_n(\varepsilon) = \frac{c_n(\varepsilon) \pm i s_n(\varepsilon)}{c_{n-1}(\varepsilon) \pm i s_{n-1}(\varepsilon)},$$

$$\Delta^\pm_N(\varepsilon) = 1 + g_{N-1,N-1}(\varepsilon) J_{N-1,N}(\varepsilon) R^\pm_N(\varepsilon),$$

and $\{\Gamma_{n\mu}\}_{n=0}^{N-1}$ is the normalized eigenvector of $\bar{H}$ associated with the eigenvalue $\varepsilon_\mu$. Now, since $g_{00}(\varepsilon)$ and $J_{N-1,N}(\varepsilon)$ are real for real energies, Broad's proposal for the interpolation function $\zeta(x)$ now reads as

$$\tan\left[\pi \zeta^{-1}(\varepsilon)\right] = \frac{\text{Im}\left[G^{(+)}_{00}(\varepsilon) - g_{00}(\varepsilon)\right]}{\text{Re}\left[G^{(+)}_{00}(\varepsilon) - g_{00}(\varepsilon)\right]} = \frac{\text{Im}\left[R^+_N(\varepsilon)/\Delta^+_N(\varepsilon)\right]}{\text{Re}\left[R^+_N(\varepsilon)/\Delta^+_N(\varepsilon)\right]} \qquad (12)$$
$$= \tan\left[\arg\left(R^+_N(\varepsilon)/\Delta^+_N(\varepsilon)\right)\right]$$

We note three points about this expression. First, it too contains the Heller's condition of equation (2). Second, it requires information related only to the finite Hamiltonian matrix $\bar{H}$ and the reference Hamiltonian $H^0$ all of which are accurately available. Moreover, this expression suggests the way to implement effectively an extension to the Heller's derivative rule. It is seen that the set $\{\tilde{\varepsilon}_\sigma\}_{\sigma=0}^{N-2}$ of zeros of the function $g_{N-1,N-1}(\varepsilon)$ is very special. These $(N-1)$ energies are precisely the eigenvalues of the matrix abbreviated from $\bar{H}$ by deleting the last row and last column. We may now arrange the two interweaving sets of energies $\{\varepsilon_\mu\}_{\mu=0}^{N-1}$ and $\{\tilde{\varepsilon}_\sigma\}_{\sigma=0}^{N-2}$ to state the interpolation conditions as

$$\zeta^{-1}(\tilde{\varepsilon}_\sigma) = \sigma + \frac{1}{\pi} \arg\left[R^+_N(\tilde{\varepsilon}_\sigma)\right]; \qquad \sigma = 0, 1, ..., N-2. \qquad (13)$$

This allows us to interpolate an accurately known $\zeta^{-1}(\varepsilon)$ at $(2N-1)$ for more accurate determination of $\{\omega^d_\mu\}_{\mu=0}^{N-1}$ via Eq. (3). This is essentially the Broad extension albeit with different set of points.

The third point is the realization that we really do not even need to interpolate the function $\zeta^{-1}(\varepsilon)$ artificially since expression (12) is actually valid for all energies, not just at a finite number of them. Broad's extension embody a natural interpolation scheme which can be implements directly. We evaluate $\zeta^{-1}(\varepsilon)$ directly and then affect an exact realization of the ansatz (3) without the need for any interpolation. This can be done by first writing $\zeta^{-1}(\varepsilon) = x$ and hence $\zeta(x) = \varepsilon$ such that $\zeta(\mu) = \varepsilon_\mu$ and noting that

$$\omega_\mu^d = \left(\frac{d\zeta}{dx}\right)_{x=\mu} = \left(\frac{d\varepsilon}{dx}\right)_{\varepsilon=\varepsilon_\mu}. \qquad (14)$$

Now, since

$$\frac{R_N^+(\varepsilon)}{\Delta_N^+(\varepsilon)} = \frac{R_N^+(\varepsilon) + g_{N-1,N-1}(\varepsilon)J_{N-1,N}(\varepsilon)\left|R_N^+(\varepsilon)\right|^2}{\left|\Delta_N^+(\varepsilon)\right|^2},$$

We can write

$$\tan \pi x = \mathrm{Im}\left[R_N^+(\varepsilon)\right] \Big/ \mathrm{Re}\left[R_N^+(\varepsilon) + g_{N-1,N-1}(\varepsilon)J_{N-1,N}(\varepsilon)\left|R_N^+(\varepsilon)\right|^2\right]. \qquad (15)$$

Differentiating both sides with respect to the energy $\varepsilon$ and taking the limit of the result as $\varepsilon$ approaches $\varepsilon_\mu$, we obtain

$$\left(\frac{dx}{d\varepsilon}\right)_{\varepsilon=\varepsilon_\mu} = -\mathrm{Im}\left[R_N^+(\varepsilon_\mu)\right] \Big/ \pi\,\Gamma_{N-1,\mu}^2 J_{N-1,N}(\varepsilon_\mu)\left|R_N^+(\varepsilon_\mu)\right|^2.$$

This yields the final result that

$$\omega_\mu^d = \frac{\pi\,\Gamma_{N-1,\mu}^2 J_{N-1,N}(\varepsilon_\mu)}{\mathrm{Im}\left[1/R_N^+(\varepsilon_\mu)\right]}. \qquad (16)$$

Without any need for interpolation.

To test the proposed extension and also the exact realization, we consider a model problem where the exact answer is known. The chosen model is a modified version of the Chebychev system [9,11], namely the one associated with the following infinite tridiagonal symmetric matrix

$$H = \begin{pmatrix} A & B & & & & & \\ B & 0 & \tfrac{1}{2} & & & \text{\Large 0} & \\ & \tfrac{1}{2} & 0 & \tfrac{1}{2} & & & \\ & & \tfrac{1}{2} & 0 & \times & & \\ & & & \times & \times & \times & \\ & \text{\Large 0} & & & \times & \times & \times \\ & & & & & \times & \times \end{pmatrix} \qquad (17)$$

where $A$ and $B$ are real constant parameters and $B \neq 0$, and this Hamiltonian is associated with the reference Hamiltonian $H_{nm}^0 = \tfrac{1}{2}(\delta_{n,m+1} + \delta_{n,m-1})$. The finite matrix $\bar{H}$ is obtained by truncating the above matrix to a finite $N \times N$ sub-matrix and the corresponding discretized continuum lies within the interval $\varepsilon \in [-1,+1]$. The exact Green's function, density function and J-matrix coefficient associated with this model are:

$$G_{00}(z) = \left[A + (2B^2 - 1)z - 2B^2\sqrt{z^2 - 1}\right]^{-1}, \qquad (18)$$

$$\rho(x) = \frac{(2B^2/\pi)\sqrt{1-x^2}}{4B^4 + (A-x)\left[A + (4B^2 - 1)x\right]}, \text{ and} \qquad (19)$$

$$R_N^\pm(x) = \left(x \pm i\sqrt{1-x^2}\right)^{-1}, \qquad (20)$$

respectively. Table I shows the results of calculating the derivative weights using the four different methods for $A = B = \frac{1}{3}$ and $N = 10$:

1. The original Heller's rule represented by Eq. (3).
2. Broad's extension of the rule where Eq. (2) is augmented by Eq. (8) and using the exact Green's function (18).
3. The J-matrix extension where Eq. (2) is augmented by Eq. (13) and using the J-matrix coefficient (20).
4. The exact J-matrix realization of the rule as stipulated by Eq. (16).

For the sake of comparing the accuracy of the four methods, the exact values are also shown. These are obtained using the exact numerical weights $\{\omega_\mu\}_{\mu=0}^{N-1}$ and density (19) in $\omega_\mu^d = \omega_\mu / \rho(\varepsilon_\mu)$. The exact J-matrix values agree with these values to machine accuracy.

The figure shows an example of the interpolation process to obtain the function $\zeta(x)$ using the J-matrix extension (method 3 above). In the figure, squares correspond to $\{\varepsilon_n\}_{n=0}^{N-1}$ and circles correspond to $\{\tilde{\varepsilon}_m\}_{m=0}^{N-2}$ whereas the thin solid curve connecting these $(2N-1)$ points represents $\zeta(x)$, which is obtained using Schlessinger fitting routine [16] of order $2N-3$. The thick solid curve represents $d\zeta/dx$ (scaled and shifted for better presentation).

The second example we give is where the Hamiltonain is the $\ell$-th partial wave kinetic energy operator $H = -\frac{1}{2}\frac{d^2}{dr^2} + \frac{\ell(\ell+1)}{2r^2}$. The matrix representation of the Hamiltonian in the orthonormal basis [2]

$$\langle r | \phi_n \rangle = a_n (\lambda r)^{\ell+1} e^{-\lambda^2 r^2/2} L_n^{\ell+1/2}(\lambda^2 r^2), \quad \text{with} \quad a_n = \sqrt{2\lambda n! / \Gamma(n+\ell+3/2)}$$

and $\lambda$ is a free parameter, has a tridiagonal representation

$$\langle \phi_n | H | \phi_m \rangle = D_{n-1} \delta_{n,m+1} + C_n \delta_{n,m} + D_n \delta_{n,m-1},$$

$$\text{with} \quad C_n = \frac{\lambda^2}{2}(2n+\ell+3/2), \text{ and } \quad D_n = \frac{\lambda^2}{2}\sqrt{(n+1)(n+\ell+3/2)}$$

This Hamilonian has been studied extensively [6,13,15]. We know that associated with it is the following solution

$$s_n(\varepsilon) = \frac{(-1)^n}{\lambda}\sqrt{\frac{\pi}{2}} a_n (2\varepsilon/\lambda^2)^{\ell+1} e^{-\varepsilon/\lambda^2} L_n^{\ell+1/2}(2\varepsilon/\lambda^2), \quad \text{and}$$

$$c_n(\varepsilon) = \frac{(-1)^n}{\lambda}\sqrt{\frac{\pi}{2}} a_n (2\varepsilon/\lambda^2)^{\ell+1} e^{-\varepsilon/\lambda^2} {}_1F_1(-n, -\frac{1}{2}-\ell; \frac{1}{2}-\ell; 2\varepsilon/\lambda^2).$$

It is also known that the energy eigenvalues of the finite $N \times N$ matrix $\bar{H}$ associated with $H$ falls as the zero of $s_N(\varepsilon)$ or equivalently the polynomial $L_N^{\ell+1/2}(2\varepsilon/\lambda^2)$. This enables use to calculate the exact value of $\omega_\mu^d$ via the quadrature formula of Eq. (5). In Table II we follow the same procedure as in the previous example but limit the comparison of results from applying the Heller derivative rule and from applying J-matrix formula Eq. (16) with the exact result for the case $\ell = 1$ and $N = 5$. As expect the J-matrix result is identical to the exact result.

In summary, we have provided an exact realization of the very powerful and simple to use Heller derivative rule. The key to this realization is the specification of the interpolation scheme as that suggested by the work of Broad. This scheme can be exactly implemented in the J-matrix context leading to an exact form for the derivative weight derivable from the properties of the finite matrix $\bar{H}$ and the Hamiltonian $H^0$.

**REFERENCES:**

[1] A. U. Hazi and H. S. Taylor, Phys. Rev. **A 2**, 1109 (1970).
[2] C. Schwartz, Ann. Phys. (N.Y.) **16**, 36 (1961).
[3] F. E. Harirs, Phys. Rev. Lett. **19**, 173 (1967).
[4] W. P. Reinhardt, D. W. Oxtoby, and T. N. Rescigno, Phys. Rev. Lett. **28**, 401 (1972).
[5] E. J. Heller, W. P. Reinhardt, and H. A. Yamani, J. Comp. Phys. **13**, 536 (1973).
[6] H. A. Yamani, M. S. Abdelmonem, J. Phys. B: Atom. Mol. **30**, 1633 (1997).
[7] Heller's rule: E. J. Heller, PhD thesis (Harvard University, 1973) unpublished.
[8] H. A. Yamani and W. P. Reinhardt, Phys. Rev. A **11**, 1144 (1975).
[9] G. Szego, *Orthogonal Polynomials* (American Mathematical Society, 1959), Vol. III, Chap. X.
[10] J. Broad, Phys. Rev. A **18**, 1012 (1978).
[11] H. A. Yamani, M. S. Abdelmonem, and A. D. Al-Haidari, Phys. Rev. A **62**, 052103 (2000).
[12] E. J. Heller, and H. A. Yamani, Phys. Rev. A **9**, 1201 (1974)
[13] H. A. Yamani, and L. Fishman, J. Math. Phys. **16**, 410 (1975).
[14] H. A. Yamani, A. D. Alhaidari, and M. S. Abdelmonem, Phys. Rev. A **64**, 042703 (2001).
[15] A. D. Alhaidari, E. J. Heller, H. A. Yamani, and M. S. Abdelmonem (Eds.), *The J-Matrix Method: Developments and Applications* (Springer, Dordrecht, Netherlands, 2008).
[16] R. W. Haymaker and L. Schlessinger, *The Pade Approximation in Theoretical Physics*, edited by G. A. Baker and J. L. Gammel (Academic Press, New York, 1970).

**TABLE CAPTION:**

**Table I**: The derivative weights $\{\omega_\mu^d\}_{\mu=0}^{N-1}$ associated with the $N \times N$ sub-matrix of (17) for $A = B = \frac{1}{3}$ and $N = 10$ and obtained using the four indicated schemes. The values in the last two columns agree to machine accuracy (not displayed).

**Table II**: The results of comparing the Heller rule Eq. (2), and the J-matrix result Eq. (16) with the exact derivative weight associated with finite ($N = 5$) energy eigenvalues of the $\ell = 1$ partial wave kinetic energy operator.

**Table I**

| $\mu$ | $\varepsilon_\mu$ | Heller Eq. (2) | Broad Eqs. (2&8) | J-matrix Eqs. (2&13) | J-matrix Eq. (16) | Exact $\omega_\mu/\rho(\varepsilon_\mu)$ |
|---|---|---|---|---|---|---|
| 0 | −0.952972 | 0.090485 | 0.093250 | 0.093250 | 0.093250 | 0.093250 |
| 1 | −0.816684 | 0.177423 | 0.176970 | 0.176970 | 0.176970 | 0.176970 |
| 2 | −0.605168 | 0.242124 | 0.242319 | 0.242319 | 0.242319 | 0.242319 |
| 3 | −0.340783 | 0.281642 | 0.281475 | 0.281475 | 0.281475 | 0.281475 |
| 4 | −0.053421 | 0.286715 | 0.286976 | 0.286976 | 0.286976 | 0.286976 |
| 5 | 0.219605 | 0.253290 | 0.252617 | 0.252617 | 0.252616 | 0.252616 |
| 6 | 0.447418 | 0.205993 | 0.207850 | 0.207848 | 0.207845 | 0.207845 |
| 7 | 0.648931 | 0.199704 | 0.196699 | 0.196695 | 0.196688 | 0.196688 |
| 8 | 0.830589 | 0.155402 | 0.159295 | 0.159300 | 0.159273 | 0.159273 |
| 9 | 0.955819 | 0.096834 | 0.085524 | 0.085916 | 0.087189 | 0.087189 |

**Table II**

| $\mu$ | $\varepsilon_\mu$ | Heller Eq. (2) | J-matrix Eq. (16) | Exact $\omega_\mu/\rho(\varepsilon_\mu)$ |
|---|---|---|---|---|
| 0 | 0.69089884 | 0.97639588 | 1.02527960 | 1.02527960 |
| 1 | 2.08912217 | 1.80175781 | 1.78939724 | 1.78939724 |
| 2 | 4.32302517 | 2.70709413 | 2.71682237 | 2.71682237 |
| 3 | 7.64230380 | 4.03613269 | 4.01574624 | 4.01574624 |
| 4 | 12.7171500 | 6.36638 | 6.50593564 | 6.50593564 |

**FIGURE CAPTION**

**Fig. 1**: The interpolation process to obtain $\zeta(x)$ associated with the quadrature matrix (17) using the J-matrix extension depicted by Eq. (2) and Eq. (13) for $N = 10$. Solid squares and circles correspond to $\{\varepsilon_n\}_{n=0}^{9}$ and $\{\tilde{\varepsilon}_m\}_{m=0}^{8}$, respectively. The thin solid curve connecting these 19 points is obtained using Schlessinger fitting routine of order 17. The thick solid curve represents $d\zeta/dx$ (scaled up by a factor of 10 and shifted down on the energy axis by 1.5 units for better presentation).

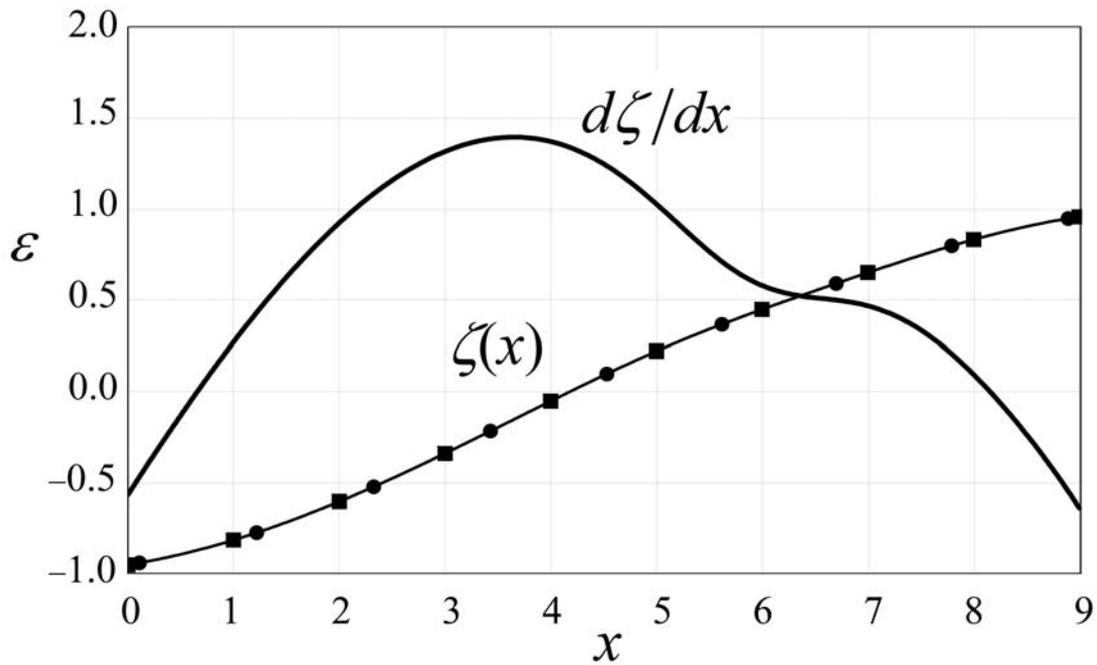

**Fig. 1**